\documentstyle[aps,amssymb,preprint]{revtex}
\tightenlines  
  
\begin{document}

\title{Asymptotic analysis of a random walk with   
a history-dependent step length}  

\author{Ronald Dickman$^{1}$\footnote{dickman@fisica.ufmg.br}, 
Francisco Fontenele Araujo Jr.$^{1}$\footnote{ffaraujo@fisica.ufmg.br} 
and Daniel ben-Avraham$^{2}$\footnote{benavraham@clarkson.edu}}
\address{$^1$Departamento de F\'{\i}sica, ICEx,   
Universidade Federal de Minas Gerais,    
Caixa Postal 702, 30.161-970,   
Belo Horizonte - MG, Brazil\\
$^2$Physics Department and Center for Statistical Physics (CISP),  
Clarkson University, Potsdam, NY 13699-5820}   

\date{\today}
 
\maketitle  
  
\begin{abstract}  
We study an unbiased, discrete-time random walk on   
the nonnegative integers, with the origin absorbing,   
and a history-dependent step length. 
Letting $y$ denote the maximum distance 
the walker has ever been from the origin,  
steps that do not change $y$ have
length $v$, while those that increase $y$   
(taking the walker to a site that has never been visited),   
have length $n$. 
The process serves as a simplified model of spreading  
in systems with an infinite number of absorbing 
configurations. Asymptotic analysis of the probability 
generating function shows that, for large $t$,  
the survival probability decays as 
$S(t) \sim t^{-\delta}$, with $\delta = v/2n$. 
Our expression for the decay exponent is 
in agreement with results obtained via numerical iteration of 
the transition matrix.
\end{abstract}  
  
\section{Introduction}  
\label{sec/intro}  
Random walks with absorbing and/or reflecting boundaries     
and/or memory serve as important models in statistical   
physics, often admitting an exact analysis. Among the   
many examples are equilibrium models for polymer adsorption   
\cite{Barber,Bell} and absorbing-state phase   
transitions \cite{Marro/Dickman}. Another motivation   
is provided by the spreading   
of an epidemic in a medium with a long memory \cite{GCR}.  
In this work we discuss a process where the   
susceptibility changes   
after the first infection and remains constant thereafter.   
  
In addition to the intrinsic interest of such an infection with         
memory, our study is motivated by the spread of activity in  
models exhibiting an infinite number of absorbing configurations,  
(INAC) typified by the pair contact process    
\cite{Jensen,Munoz/JSP}.   
Anomalies in critical spreading for INAC, such as continuously variable  
critical exponents, have been traced to a long memory in the dynamics  
of the order parameter, $\rho$, due to coupling to an auxiliary field  
that remains frozen in regions where $\rho = 0$  
\cite{Munoz/JSP,Munoz/PRL}.  
INAC appears to be particularly relevant to the transition to spatiotemporal  
chaos, as shown in a recent study of a coupled-map lattice with  
``laminar" and ``turbulent" states, which revealed continuously  
variable spreading exponents \cite{Bohr}.  
Grassberger, Chat\'e and Rousseau \cite{GCR} proposed that  
spreading in INAC could be understood by studying a model with a  
{\it unique} absorbing configuration, but in which the spreading rate   
of activity into previously inactive regions is different  
than for revisiting a region that has already been active.   
  
In light of the anomalies found in spreading in models with INAC  
or with a memory, we are interested in studying the effect of   
such a memory on the scaling behavior in a model whose asymptotic  
behavior can be determined exactly.  Of particular interest  
is the survival probability $S(t)$ (i.e., not to have fallen into  
the absorbing state up to time $t$).  
The simplest example of  
such a model is an unbiased random walk on the nonnegative  
integers, with the origin absorbing, for which $S(t) \sim t^{-\delta}$  
with $\delta = 1/2$.  It was recently shown that  
such a walker exhibits a continuously variable exponent $\delta$  
when subject to a mobile, partial reflector.  The latter is initially  
one site to the right of the walker.  Each time the walker steps  
onto the site occupied by the reflector, it is reflected one step  
to the left with probability $r$ (it remains at its new  
location with probability $1\!-\!r$); in either case, the reflector   
is pushed forward one site in this encounter.  The survival exponent  
$\delta = (1\!+\!r)/2$ in this process    
\cite{Dickman/ben-Avraham}. Since the reflector effectively 
records the {\it span} of the walk  
(i.e., the rightmost site yet visited), its interaction with the  
walker represents a memory.  

In the present work, we study a random walk with  
memory of a different form: if the target site $x$ lies 
in the region that has been visited before
(that is, if $x$ itself has been visited, or lies between
two sites that have been visited), 
then the step length is $v$; otherwise the step length is $n$. 
If $v>n$, the random walk evolves in a 
\textit{hostile} enviroment, while for $v<n$, the enviroment may be   
regarded as \textit{friendly}. On the basis of an exact  
solution for the probability generating function, we obtain the  
decay exponent $\delta$.    

The balance of this paper is organized as follows.  In Sec. \ref{sec/rw-hostile} we  
analyze the specific case of a random walk in a hostile enviroment with  
$v=2$ and $n=1$, present the solution of the generating function, and obtain 
the asymptotic behavior of the survival probability.  
In Sec. \ref{sec/arbitrary-sl} we extend the analysis to arbitrary step lengths 
$v$ and $n$ (with $v$ and $n$ natural numbers).  
In Sec. \ref{sec/numerical} we present exact numerical results  
for finite times (from iteration of the probability transfer matrix)  
that complement and extend the asymptotic analysis.  Sec. \ref{sec/discussion}  
contains a brief summary and discussion.

\section{Random walk in a hostile enviroment}  
\label{sec/rw-hostile}
\subsection{Model and Generating Function}
Consider an unbiased, discrete time random walk on the nonnegative   
integers, with the origin absorbing. We denote the position of   
the walker at time $t$ by $x_{t}$ and suppose that $x_{0}=1$.  
To define precisely the history dependence, 
let $y_t \equiv \max_t \{x_t\}$.  Then, if $x_t \leq y_t - 2$,
the walker jumps two lattice spacings to the left or the right.
If, however, $x_t = y_t$, it can move (with equal probability)
to $y_t - 2$, or to $y_t + 1$.  (In the latter case
$y_{t+1} = y_t + 1$. Notice that $y_t - x_t \geq 0$ must be even.)
Let sites 1,...,$y_t$ define the {\it known region}; 
steps to sites within the known region have length two, while those
that take the walker into the {\it unknown} region ($x > y_t$)
are of unit length. 

Evidently, the process $x_{t}$ is non-Markovian, since the   
transition probability into a given site depends on whether    
it lies in the known or the unknown region.
We can however transform the model   
to a Markov process by enlarging the state space   
\cite{van Kampen} to include   
the boundary between the two regions.  Evidently,   
the stochastic process $(x_{t}, y_{t})$ is Markovian.
The transitions (all with probability $1/2$) for the Markov chain   
are restricted to the set $E \subset {\Bbb Z}^{2}$   
specified by   
\[   
E  
=  
\{  
(x,y) \in {\Bbb Z}^{2}:  
x \geq -1, \;\;  
y \geq 1, \;\;  
x \leq y,  \;\;  
y-x \mbox{ is even}  
\}  
\]  
as represented in Fig. 1.  
 
Let $P(x,y,t)$ denote the probability of
state $(x,y)$ (for $x>0$), at time $t$.
$P(x,y,t)$ follows the evolution equation  
\begin{equation}  
\label{evolution}  
P(x,y,t+1) = \frac{1}{2}P(x+2,y,t) + \frac{1}{2}P(x-2,y,t),  
\hspace{1cm} \mbox{for $x<y$},  
\end{equation}  
with $P(1,1,t)=\delta_{0,t}$. Eq. (\ref{evolution}) is 
subject to two boundary conditions. The first is 
the absorbing condition for $x \leq 0$
\begin{equation}  
\label{bc at x=0}  
P(x,y,t) = 0, \hspace{1cm} \mbox{for $x \leq 0$.}  
\end{equation}  
The second applies along the diagonal $x\!=\!y$.    
In this case, it is convenient to define 
$D(y,t) \equiv P(y,y,t)$. On the diagonal the 
evolution equation is   
\begin{equation}  
\label{bc at x=y}  
D(y,t+1) = \frac{1}{2}\;D(y-1,t) + \frac{1}{2}P(y-2,y,t),  
\hspace{1cm} \mbox{for $y \geq 2$}.  
\end{equation}  

To solve the problem specified by Eqs. (\ref{evolution})   
- (\ref{bc at x=y}), we introduce a generating function: 
\begin{equation}  
\label{generating}  
\hat{P}(x,y,z) = \sum_{t=0}^{\infty} P(x,y,t) \; z^{t},  
\end{equation}  
where $0\leq z \leq 1$. Multiplying Eq. (\ref{evolution}) 
by $z^{t}$, summing over $t$ and shifting the sum index  
where necessary, one finds that the generating   
function satisfies   
\begin{eqnarray}  
\label{evolution/generating/x<y-4}  
\frac{1}{z}\hat{P}(x,y)   
&=&   
\frac{1}{2}\hat{P}(x+2,y)   
+   
\frac{1}{2}\hat{P}(x-2,y),  
\hspace{1cm} \mbox{for $x \leq y-4$}\\  
\label{evolution/generating/x=y-2}  
\frac{1}{z}\hat{P}(y-2,y)   
&=&   
\frac{1}{2}\hat{D}(y)   
+   
\frac{1}{2}\hat{P}(y-4,y),  
\hspace{2.1cm} \mbox{for $x=y-2$},  
\end{eqnarray}  
(we drop the argument $z$ for brevity), where    
$\hat{D}(y)$ is defined by an expression analogous   
to Eq. (\ref{generating}). The   
initial condition implies $\hat{D}(1)=1$; the   
boundary conditions are  
\begin{eqnarray}  
\label{bc at x=0/generating}  
\hat{P}(x,y) & = & 0, \hspace{6cm} \mbox{for $x \leq 0$},\\   
\label{bc at x=y/generating}  
\frac{1}{z}\hat{D}(y)   
& = &   
\frac{1}{2}\hat{D}(y-1) + \frac{1}{2}\hat{P}(y-2,y),  
\hspace{1.5cm} \mbox{for $y \geq 2$}.  
\end{eqnarray}
  
Next, we focus on Eq. (\ref{bc at x=y/generating})   
in order to eliminate $\hat{D}(y)$ in   
Eq. (\ref{evolution/generating/x=y-2}), and then find   
a recurrence relation for $\hat{P}(x,y)$.  
  
Substituting Eq. (\ref{evolution/generating/x=y-2}) in   
Eq. (\ref{bc at x=y/generating}) we have   
\begin{equation}  
\label{rr1}  
\left[  
1-\left(\frac{z}{2}\right)^{2}  
\right]  
\;\hat{D}(y)  
=  
\frac{z}{2}\; \hat{D}(y-1)  
+  
\left( \frac{z}{2} \right)^{2} \;  
\hat{P}(y-4,y).  
\end{equation}  
If we subtract $z/2$ times Eq. (\ref{rr1}), evaluated   
at $y-1$,   
from the corresponding equation for $y$, we find  
\begin{eqnarray*}  
\left[  
1-\left( \frac{z}{2}\right)^{2}  
\right] \;  
\left[  
\hat{D}(y)-\frac{z}{2}\;\hat{D}(y-1)  
\right]   
&=&  
\frac{z}{2}  
\left[  
\hat{D}(y-1)-\frac{z}{2}\;\hat{D}(y-2)  
\right]+\\  
\;\;\;&+&  
\left( \frac{z}{2}\right)^{2}  
\left[  
\hat{P}(y-4,y)-\frac{z}{2}\hat{P}(y-5,y-1)  
\right].   
\end{eqnarray*}  
Using Eq. (\ref{bc at x=y/generating}) we eliminate   
$\hat{D}(y)$, $\hat{D}(y-1)$ and $\hat{D}(y-2)$ to obtain   
\begin{eqnarray}  
\nonumber  
\frac{z}{2}  
\left[  
1-\left( \frac{z}{2}\right)^{2}  
\right] \;  
\hat{P}(y-2,y)   
&=&  
\left(\frac{z}{2}\right)^{2}  
\hat{P}(y-3,y-1)  
+\\  
\nonumber  
&&\;+\left( \frac{z}{2}\right)^{2}  
\left[  
\hat{P}(y-4,y)-\frac{z}{2}\hat{P}(y-5,y-1)  
\right],  
\end{eqnarray}  
yielding the recurrence relation    
\begin{equation}  
\label{PP-relation}  
\left[  
\frac{4-z^{2}}{2z}  
\right] \;  
\hat{P}(y-2,y)  
-  
\hat{P}(y-4,y)  
=  
\hat{P}(y-3,y-1) -   
\frac{z}{2}\hat{P}(y-5,y-1).  
\end{equation}

Eq. (\ref{evolution/generating/x<y-4}) relates   
$\hat{P}(x,y)$ for different $x$, at fixed $y$.   
We therefore impose separation of variables and write,   
for $x \leq y-2$,  
\begin{equation}  
\label{product}  
\hat{P}(x,y)   
=  
\hat{A}(x) \; \hat{B}(y). 
\end{equation}   
Eq. (\ref{bc at x=0}) requires $\hat{A}(x) = 0$ 
for $x \leq 0$, which is satisfied if
\begin{equation}  
\label{A/21}
\hat{A}(x)   
=  
\left \{  
\begin{array}{ll}  
\lambda^{x} - \lambda^{-x} & \mbox{, for $x \geq 0$,}\\  
0 & \mbox{, for $x < 0$.}  
\end{array}  
\right.  
\end{equation}   
In this context, Eq. (\ref{evolution/generating/x<y-4}) implies:   
\begin{equation}  
\label{lambda}  
\lambda 
= 
\left(
\frac{1}{z} + \sqrt{\frac{1}{z^{2}} -1}
\right)^{1/2}.  
\end{equation}    
(Note that use of the second solution,   
$\lambda_- = (z^{-1} - \sqrt{z^{-2} -1})^{1/2} = \lambda^{-1}$, would simply  
result in a change in the sign of  
$\hat{A}$.)  
Substituting Eq. (\ref{product}) in   
Eq. (\ref{PP-relation}) we find   
\begin{equation}  
\label{B/ratio/21}  
\frac{\hat{B}(y)}{\hat{B}(y-1)}  
=  
\frac{2z\;\hat{A}(y-3)-z^{2}\;\hat{A}(y-5)}  
{(4-z^{2})\;\hat{A}(y-2)-2z\;\hat{A}(y-4)}.  
\end{equation}  
\subsection{The survival probability}  
The survival probability is   
\[  
S(t)  
=  
\sum_{y=2}^{\infty}  
\sum_{k=0}^{\left[ \frac{y-1}{2}\right]}  
P(y-2k,y,t),
\]  
where $[\;]$ denotes the integer part of its argument. 
The corresponding generating function is:   
\begin{eqnarray}  
\nonumber  
\hat{S}(z)  
&=&  
\sum_{t=0}^{\infty}  
S(t) \; z^{t}\\  
&=&  
\label{S}  
\hat{S}_{P}(z)   
+ \hat{S}_{D}(z),  
\end{eqnarray}  
where   
$  
\hat{S}_{P}(z)   
=  
\sum_{y=2}^{\infty} 
\sum_{k=1}^{[(y-1)/2]} 
\hat{P}(y-2k,y)   
$   
and  
$  
\hat{S}_{D}(z)  
=  
\sum_{y=2}^{\infty} \hat{D}(y)  
$.   
We study these series separately. To begin, 
we insert Eq.(\ref{product}) in $\hat{S}_{P}$ 
to obtain   
\begin{equation}
\label{S/P/21}
\hat{S}_{P} (z)  
=  
\sum_{y=2}^{\infty}  
\sum_{k=1}^{\left[\frac{y-1}{2} \right]} 
\hat{A}(y-2k)   
\;
\hat{B}(y).  
\end{equation}  
Next we examine $\hat{S}_{D}(z)$.  
Iterating Eq. (\ref{bc at x=y/generating}), we have     
\begin{equation}  
\label{DP-relation}  
\hat{D}(y)   
=   
\left( \frac{z}{2}\right)^{y-1} +   
\sum_{j=2}^{y}  
\left( \frac{z}{2}\right)^{y+1-j}  
\hat{P}(j-2,j).  
\end{equation}  
Summing Eq. (\ref{DP-relation}) over $y \geq 2$   
we find   
\begin{equation}  
\label{S/D/21}  
\hat{S}_{D}(z)  
=  
\frac{z}{2-z}  
\left[ 
1  
+  
\sum_{y=2}^{\infty}  
\hat{A}(y-2)\;  
\hat{B}(y)  
\right].  
\end{equation}  
\subsection{Asymptotic analysis}  
We address the \textit{Tauberian} problem   
\cite{Montroll/1965}  
of extracting the large-$t$   
asymptotics of $S(t)$ from the dominant 
singularity of its generating function $\hat{S}(z)$,   
as $z \uparrow 1$.    
In order to study this limit,   
let $z = 1 - \epsilon$, with  
$\epsilon \downarrow 0$.  
We will show that as $t \to \infty$,   
the dominant contribution to the survival probability comes   
from $\hat{S}_{P}(z)$.    

To determine the asymptotic behavior of $\hat{S}_P$, 
we analyze $\hat{B}(y)$ and the sum $\sum_{k=1}^{[(y-1)/2]} \hat{A}(y-2k)$ separately.
First, we focus on $\hat{B}$; in light of Eq. (\ref{B/ratio/21}), it is convenient 
to write, 
\begin{equation}
\label{B/product/21}
\hat{B}(y)
=
\hat{B}(3)
\prod_{k=4}^{y}
\frac{\hat{B}(k)}{\hat{B}(k-1)}.
\end{equation}
>From Eq. (\ref{product}) we have   
$
\hat{B}(3)   
=   
\hat{P}(1,3)\; / \;\hat{A}(1)
$, 
with 
$\hat{A}(1) \simeq \sqrt{2\epsilon}$, as   
$\epsilon \downarrow 0$. 
(We use the symbol ``$\simeq$'' to indicate asymptotic 
equality as $\epsilon \downarrow 0$.) 
On the other hand,  
Eq. (\ref{bc at x=y/generating}) implies that  
$  
\hat{P}(1,3)  
=  
\frac{2}{z} \; \hat{D}(3) -\hat{D}(2),  
$   
where $\hat{D}(2) = z/2$. Iterating Eq. (\ref{bc at x=y})   
we have  
\[  
D(3,t)  
=  
\left \{  
\begin{array}{rl}  
\frac{1}{2^{t}} & \mbox{, if $t$ is odd}\\  
0 & \mbox{, if $t$ is even}  
\end{array}  
\right.  
\]  
so that 
$ 
\hat{D}(3)   
=  
2z\;/\;(4-z^{2})  
$.   
Evidently, $\hat{P}(1,3) \simeq \frac{1}{6}$, as   
$\epsilon \downarrow 0$, and therefore   
\begin{equation}
\label{B(3)/21}
\hat{B}(3) \simeq \frac{1}{6\sqrt{2\epsilon}}.
\end{equation}
The ratio $\hat{B}(k) / \hat{B}(k-1)$ may be    
analyzed by inserting Eq. (\ref{A/21}) in Eq. (\ref{B/ratio/21})  
\[
\frac{\hat{B}(k)}{\hat{B}(k-1)}
=
\frac{\lambda^{k}(2z\lambda^{-3}-z^{2}\lambda^{-5}) - 
\lambda^{-k}(2z\lambda^{3}-z^{2}\lambda^{5})}
{\lambda^{k}[(4-z^{2})\lambda^{-2}-2z\lambda^{-4}] - 
\lambda^{-k}[(4-z^{2})\lambda^{2}-2z\lambda^{4}]}
\]
For small $\epsilon$,  
$\lambda = 1 + \frac{1}{2}\sqrt{2\epsilon} + {\cal O}(\epsilon)$. 
Then,
\[
\frac{\hat{B}(k)}{\hat{B}(k-1)}
\simeq
\frac{\lambda^{k} - \lambda^{-k}
-\frac{1}{2}\sqrt{2\epsilon}(\lambda^{k} + \lambda^{-k})}
{\lambda^{k} - \lambda^{-k} + 
\sqrt{2\epsilon}(\lambda^{k} + \lambda^{-k})}.
\]
Now, letting $\Lambda = \ln(\lambda)$, we have 

\begin{eqnarray*}
\frac{\hat{B}(k)}{\hat{B}(k-1)}
&\simeq&
\frac{\tanh{\left(\frac{\sqrt{2\epsilon}}{2} \, k\right)}
-\frac{1}{2}\sqrt{2\epsilon}}
{\tanh{\left(\frac{\sqrt{2\epsilon}}{2} \, k\right)}
+\sqrt{2\epsilon}}.
\end{eqnarray*}
Letting $\phi_{k}=\tanh\left( \frac{\sqrt{2\epsilon}}{2}k\right)$, this yields,
\[
\sum_{k=4}^y
\ln 
\left[
\frac{\hat{B}(k)}{\hat{B}(k-1)}
\right]
\simeq
- \left( \frac{3}{2}\right) \sqrt{2\epsilon}
\sum_{k=4}^{y}
\frac{1}{\phi_{k}}.
\]
Approximating the sum by an integral, we have
\begin{eqnarray*}
\sum_{k=4}^{y}
\frac{1}{\phi_{k}}
&\simeq&
\frac{2}{\sqrt{2\epsilon}}
\int_{2\sqrt{2\epsilon}}^{\frac{\sqrt{2\epsilon}}{2}y}
\frac{1}{\tanh(w)}\;dw\\
&\simeq&
\frac{2}{\sqrt{2\epsilon}}
\ln
\left[
\frac{\sinh \left( \frac{\sqrt{2\epsilon}}{2} y \right)}
{\sinh \left( 2\sqrt{2\epsilon} \right)}
\right],
\end{eqnarray*}
and hence,
\begin{equation}
\label{product/ratio/21}
\prod_{k=4}^{y}
\frac{\hat{B}(k)}{\hat{B}(k-1)}
\simeq
C \;
\frac{\epsilon^{\frac{3}{2}}}{\sinh^{3} \left[ \frac{\sqrt{2\epsilon}}{2}y\right]},
\end{equation}
where $C$ is a constant. Inserting Eqs. (\ref{B(3)/21}) and 
(\ref{product/ratio/21}) in Eq. (\ref{B/product/21}) we find, 
\begin{equation}
\label{B/asymp/21}
\hat{B}(y)
\sim
\frac{\epsilon}{\sinh^{3} \left( \frac{\sqrt{2\epsilon}}{2}y\right)}.
\end{equation}
(By ``$\sim$'' we mean asymptotic proportionality as $\epsilon \downarrow 0$, 
i.e., multiplicative constants are ignored.) 
Finally, we note that as  $\epsilon \downarrow 0$, 
\begin{equation}
\label{sum/A/asymp/21}
\sum_{k=1}^{\left[\frac{y-1}{2} \right]}
\hat{A}(y-2k)
\simeq
\frac{\left[ 
\hat{A} 
\left( 
\frac{y}{2} \right) 
\right]^{2}}
{2\sqrt{2\epsilon}}
\sim
\epsilon^{-\frac{1}{2}}
\sinh^{2}
\left(
\frac{\sqrt{2\epsilon}}{4} y
\right).
\end{equation}
With these results, we are in a position to analyze the asymptotic 
behavior of $\hat{S}_{P}$ and $\hat{S}_{D}$.

First, we determine the asymptotic behavior of $\hat{S}_{P}(z)$.   
Substituting Eqs. (\ref{sum/A/asymp/21}) and (\ref{B/asymp/21}) in 
Eq. (\ref{S/P/21}) we 
find
\[
\hat{S}_{P} 
\sim
\epsilon^{\frac{1}{2}}
\sum_{y=2}^{\infty}
\frac{\sinh^{2} \left( \frac{\sqrt{2\epsilon}}{4} y \right)}
{\sinh^{3} \left( \frac{\sqrt{2\epsilon}}{2} y\right)}.
\]
Denoting the sum by $H(\epsilon)$, we have
\[
H(\epsilon)
\sim
\epsilon^{-1/2}
\int_{\sqrt{2\epsilon}}^{\infty}
\frac{\sinh^{2}\left(\frac{1}{2}w\right)}{\sinh^{3}(w)}\;dw.
\]
Let us denote the 
integrand by $f(w)$ and the integral by $I(\epsilon)$. 
Since $f(w)$ has a pole of order $1$, we introduce the Laurent expansion 
\[
f(w) 
= 
\frac{1}{4} \; w^{-1}
+
\sum_{k=0}^{\infty}
a_{k}\; w^{k}
\]
and integrate the series term by term. Noting that the dominant 
contribution, as $\epsilon \downarrow 0$, comes from the first term, 
we have 
\[
I(\epsilon)
\simeq
\frac{1}{4}
\int_{\sqrt{2\epsilon}}^{1} 
w^{-1}\;dw
\simeq
-\frac{1}{8} \; \ln(\epsilon).
\]
Thus,
\[
\hat{S}_{P} 
\sim
-
\ln(\epsilon).
\]
Using the same line of reasoning, it can be shown  
$\lim_{\epsilon \downarrow 0} \hat{S}_{D} / \hat{S}_{P} = 0$.
Therefore, the dominant singular behavior of $\hat{S}$ as $z \uparrow 1$ is 
given by:
\[
\hat{S}(z)  
\sim  
-
\ln(1-z).
\]   
The coefficient of $z^{t}$ in the expansion of 
$-\ln(1-z)$ is $t^{-1}$, and so the 
survival probability decays asymptotically as $t^{-1}$.

\section{Arbitrary step lengths}
\label{sec/arbitrary-sl}
In this section, we generalize the analysis of Sec. \ref{sec/rw-hostile}  
to a walker with an arbitrary history-dependent step length. 
Let $v$ be the step length for target sites
in the known region,  
and $n$ the step length in case the target site 
lies in the unknown region. 
We consider the Markov chain $(x_{t}, y_{t})$, 
with $y_{t}$ as defined in Sec. \ref{sec/rw-hostile} and 
transitions with probability $1/2$. 
The probability $P(x,y,t)$ follows the equation   
\begin{equation}
\label{evolution/vn}  
P(x,y,t+1) = \frac{1}{2}P(x+v,y,t) + \frac{1}{2}P(x-v,y,t),  
\hspace{1cm} \mbox{for $x<y$}, 
\end{equation}   
with $P(1,1,t)=\delta_{t,0}$. Eq. (\ref{evolution/vn}) 
is subject to two boundary conditions, the first, Eq. (\ref{bc at x=0}), is
due to the absorbing condition.
The second applies along the diagonal $x=y$.  
Defining $D(y,t) = P(y,y,t)$, as before, we have   
\begin{equation}
\label{bc at x=y/vn}  
D(y,t+1) = \frac{1}{2}\;D(y-n,t) + \frac{1}{2}P(y-v,y,t),  
\hspace{1cm} \mbox{for $y \geq n+1$}.  
\end{equation}
Introducing the generating functions $\hat{P}(x,y)$ and $\hat{D}(y)$   
as in Sec. \ref{sec/rw-hostile}, one readily finds, 
\begin{eqnarray}   
\label{evolution/generating/x<y-v}
\frac{1}{z}\hat{P}(x,y)   
&=&   
\frac{1}{2}\hat{P}(x+v,y)  
+  
\frac{1}{2}\hat{P}(x-v,y),  
\hspace{0.5cm} \mbox{for $x \leq y-2v$,}\\   
\frac{1}{z}\hat{P}(y-v,y)   
&=&   
\frac{1}{2}\hat{D}(y)   
+    
\frac{1}{2}\hat{P}(y-2v,y),   
\hspace{1.6cm} \mbox{for $x=y-v$.}   
\end{eqnarray}   
The initial condition is $\hat{D}(1)=1$, and the   
boundary conditions are   
\begin{eqnarray}  
\label{bc at x=0/vn}
\hat{P}(x,y) & = & 0, \hspace{4.9cm} \mbox{ for $x \leq 0$} \\  
\label{bc at x=y/generating/vn}
\frac{1}{z}\hat{D}(y)  
& = &  
\frac{1}{2}\hat{D}(y-n)  
+  
\frac{1}{2}\hat{P}(y-v,y), 
\hspace{0.5cm}  
\mbox{for $y \geq n+1$}.  
\end{eqnarray}  
Proceeding as in Sec. \ref{sec/rw-hostile}, one finds the 
recurrence relation:  
\begin{equation}  
\label{PP-relation/vn}  
\left[   
\frac{4-z^{2}}{2z}  
\right] \;  
\hat{P}(y-v,y)  
-  
\hat{P}(y-2v,y)  
=  
\hat{P}(y-v-n,y-n)  
-  
\frac{z}{2}\hat{P}(y-2v-n,y-n).  
\end{equation}  
The solution for $\hat{P}(x,y)$ is 
again of the form of Eq. (\ref{product}), with $\hat{A}$ again given by
Eq. (\ref{A/21}), but with,
\[
\lambda
=
\left(
\frac{1}{z}
+
\sqrt{\frac{1}{z^{2}} -1}
\right)^{1/v}
\]
With this, one readily finds,
\begin{equation}  
\label{B/ratio/vn}  
\frac{\hat{B}(y)}{\hat{B}(y-n)}  
=  
\frac{2z\;\hat{A}(y-v-n)-z^{2}\;\hat{A}(y-2v-n)}  
{(4-z^{2})\;\hat{A}(y-v)-2z\;\hat{A}(y-2v)}.  
\end{equation}  
Since $y=nj+1$ and $x = y - vk = nj - vk + 1$, 
$\hat{S}(z)$ is given by
\begin{equation}
\hat{S}(z)
=
\sum_{j=1}^{\infty} 
\sum_{k=0}^{L}
\hat{P}(nj - vk + 1, \; nj + 1),
\end{equation}
where $L = \left[ \frac{nj}{v}\right]$.
We define
\begin{eqnarray}
\nonumber
\hat{S}_{P}(z)
&=&
\sum_{j=1}^{\infty} 
\sum_{k=1}^{L}
\hat{P}(nj-vk+1, \; nj+1)\\
&=&
\label{S/P/vn}
\sum_{j=1}^{\infty} 
\sum_{k=1}^{L}
\hat{A}(nj-vk+1) \;
\hat{B}(nj+1)
\end{eqnarray}
and 
\[
\hat{S}_{D}(z)
=
\sum_{j=1}^{\infty} 
\hat{D}(nj+1).
\]
Iterating Eq. (\ref{bc at x=y/generating/vn}) we find  
\begin{equation}  
\label{DP-relation/vn}  
\hat{D}(nj+1)   
=   
\left( \frac{z}{2}\right)^{j} +   
\sum_{k=1}^{j}  
\left( \frac{z}{2}\right)^{j+1-k}  
\hat{P}(nk-v+1,nk+1).  
\end{equation}  
Summing Eq. (\ref{DP-relation/vn}) over $y \geq n+1$,  
and inserting the expressions found previously for $\hat{P}(x,y)$, 
we have
\begin{equation}  
\label{S/D/vn}  
\hat{S}_{D}(z)  
=  
\frac{z}{2-z}  
\left[  
1  
+  
\sum_{j=1}^{\infty}  
\hat{A}(nj-v+1)\;\hat{B}(nj+1)  
\right].  
\end{equation}

In order to determine the asymptotic behavior of Eqs. (\ref{S/P/vn}) and 
(\ref{S/D/vn}), we analyze $\hat{A}$, $\hat{B}$ and the sum  
$\sum_{k=1}^{L} \hat{A}(nj-vk+1)$ separately.
First, we write $\hat{B}$ in the form: 
\begin{equation}
\label{B/product/vn}
\hat{B}(nj+1)
=
\hat{B}(nk_{0} +1)
\prod_{k=k_{0}+1}^{j}
\frac{\hat{B}(nk+1)}{\hat{B}(nk-n+1)},
\end{equation}
where $k_{0}$ is the smallest positive integer such that the argument of $\hat{A}(x)$ 
is positive. Note that 
\begin{equation}
\label{B(1-nk0)}
\hat{B}(nk_{0}+1)
\simeq 
C_{1}(v,n,k_0) \; \epsilon^{-\frac{1}{2}},
\end{equation} 
with $C_{1}(v,n,k_0)$ a coefficient which depends on $v$, $n$ and $k_0$. 
The ratio $\hat{B}(y) / \hat{B}(y-n)$ may be written,  
\[
\frac{\hat{B}(y)}{\hat{B}(y-n)}
=
\frac{\lambda^{y}(2z\lambda^{-v-n}-z^{2}\lambda^{-2v-n}) - 
\lambda^{-y}(2z\lambda^{v+n}-z^{2}\lambda^{2v+n})}
{\lambda^{y}[(4-z^{2})\lambda^{-v}-2z\lambda^{-2v}] - 
\lambda^{-y}[(4-z^{2})\lambda^{v}-2z\lambda^{2v}]}.
\]
For small $\epsilon$,
$\lambda \simeq 1 + \frac{1}{v}\sqrt{2\epsilon} $, and 
\begin{eqnarray*}
\frac{\hat{B}(y)}{\hat{B}(y-n)}
&\simeq&
\frac{\lambda^{y} - \lambda^{-y}
-\frac{n}{v}\sqrt{2\epsilon}(\lambda^{y} + \lambda^{-y})}
{\lambda^{y} - \lambda^{-y} + 
\sqrt{2\epsilon}(\lambda^{y} + \lambda^{-y})}.
\end{eqnarray*}
As before, let $\Lambda = \ln(\lambda)$.  Then,  
\begin{eqnarray*}
\frac{\hat{B}(y)}{\hat{B}(y-n)}
&\simeq&
\frac{\tanh{\left(\frac{\sqrt{2\epsilon}}{v} \, y\right)}
-\frac{n}{v}\sqrt{2\epsilon}}
{\tanh{\left(\frac{\sqrt{2\epsilon}}{v} \, y\right)}
+\sqrt{2\epsilon}}.
\end{eqnarray*}
A calculation analogous to that leading to Eq. (\ref{B/asymp/21})
then yields,
\begin{equation}
\label{B/asymp}
\hat{B}(nj+1)
\simeq
C_2(v,n,k_0)\;
\frac{\epsilon^{\frac{v}{2n}}}{\sinh^{1+\frac{v}{n}} \left[ \frac{\sqrt{2\epsilon}}{v}(1+nj)\right]},
\end{equation}
with $C_2(v,n,k_0)$ a coefficient which depends on $v$, $n$ and $k_0$. 
Finally, we note that as  $\epsilon \downarrow 0$, 
\begin{equation}
\label{sum/A/asymp}
\sum_{k=1}^{L}
\hat{A}(nj-vk+1)
\simeq
\frac{\left[ 
\hat{A} 
\left( 
\frac{nj+1}{2} \right) 
\right]^{2}}
{v\sqrt{2\epsilon}}
\sim
\epsilon^{-\frac{1}{2}}
\sinh^{2}
\left[
\frac{\sqrt{2\epsilon}}{2v} (nj+1)
\right].
\end{equation}
With these results, we are in a position to analyze the asymptotic 
behavior of $\hat{S}_{P}$ and $\hat{S}_{D}$.

First, we determine the asymptotic behavior of $\hat{S}_{P}(z)$.   
Substituting Eqs. (\ref{sum/A/asymp}) and (\ref{B/asymp}) in Eq. (\ref{S/P/vn}) we 
find
\[
\hat{S}_{P} 
\sim
\epsilon^{\frac{v}{2n} - \frac{1}{2}}
\sum_{j=1}^{\infty}
\frac{\sinh^{2} \left[ \frac{\sqrt{2\epsilon}}{2v} (nj+1) \right]}
{\sinh^{1+\frac{v}{n}} \left[ \frac{\sqrt{2\epsilon}}{v} (nj+1)\right]}.
\]
As before, we approximate the sum, denoted $H(\epsilon)$, by an integral,
\[
H(\epsilon)
\sim
\frac{v}{n\sqrt{2}}\;\epsilon^{-1/2}
\int_{q\epsilon^{1/2}}^{\infty}
\frac{\sinh^{2}\left(\frac{1}{2}w\right)}{\sinh^{1+\frac{v}{n}}(w)}\;dw,
\]
with $q=\frac{(1+n)\sqrt{2}}{v}$. Let us denote the 
integrand by $f(w)$ and the integral by $I(\epsilon)$. First, 
note that if $v<n$, then $f(w)$ is bounded, and $I(\epsilon)$ converges. Therefore,
\[
\hat{S}_{P}
\sim
\epsilon^{\frac{v}{2n}-1}, 
\hspace{1cm} \mbox{for $v<n$.}
\]
On the other hand, if $v>n$, then $f(w)$ diverges as $w\downarrow 0$ and decays 
exponentially for $w \gg 1$. 
In particular, if $\frac{v}{n} = 2m$, with $m=1,2,3, \ldots$, then $f(w)$ 
has a pole of order $m$. Introducing the Laurent expansion 
\[
f(w) 
= 
\sum_{k=0}^{\infty}
a_{k}(m)\; w^{k}
+
\sum_{k=1}^{m} 
b_{k}(m) \; w^{-k}
\]
and integrating the series term by term, we note that the dominant contribution to $\hat{S}_P$,
as $\epsilon \downarrow 0$, comes from the term proportional 
to $w^{-1}$. Thus, 
\[
I(\epsilon)
\sim
b_{1}(m)
\int_{q\epsilon^{1/2}}^{1} 
w^{-1}\;dw
\sim
-b_{1}(m) \; \ln(\epsilon).
\]
Since $b_{1}(m)$ is the residue of $f(w)$ at $w=0$, we may relate it to an  
integral around a closed contour containing the origin in the complex-$w$ plane. 
In this manner we can establish the recurrence relation:  
\[
b_{1}(m)
=
-\frac{1}{2m}
\left[
b_{1}(m-1) 
-
b_{1}(m-2)
+
b_{1}(m-3)
-
\ldots
+
(-1)^{m}\,3b_{1}(1)  
\right],
\]
where $b_{1}(1) = 1/4$, as found in Sec. \ref{sec/rw-hostile}. Observe that the $b_{1}(m)$ 
alternate in sign. Thus,
\[
\hat{S}_{P} 
\sim
(-1)^{\frac{v}{2n}}\;
\epsilon^{\frac{v}{2n}-1}\;
\ln(\epsilon), 
\hspace{1cm} \mbox{for $v=2m\,n$.}
\]
Using the same line of reasoning, it can be shown that if $n < v \neq 2m\,n$, 
then $\hat{S}_{P} \sim \epsilon^{\frac{v}{2n}-1}$. Moreover, 
$\lim_{\epsilon \downarrow 0} \hat{S}_{D} / \hat{S}_{P} = 0$.
Therefore, the dominant singular behavior of $\hat{S}$ as $z \uparrow 1$ is 
given by:
\[
\hat{S}(z)  
\sim  
\left \{
\begin{array}{ll}
(-1)^{\frac{v}{2n}}\;(1-z)^{\frac{v}{2n}-1}\;\ln(1-z)\;, &\mbox{for $v=2n, 4n, 6n, \ldots$,}\\
&\\
(1-z)^{\frac{v}{2n}-1}\;, & \mbox{otherwise.}
\end{array}
\right.
\]   
The coefficient of $z^{t}$ (for large $t$) in the expansion of 
$(-1)^{\frac{v}{2n}}(1-z)^{\frac{v}{2n}-1}\;\ln(1-z)$ is proportional to 
$t^{-\frac{v}{2n}}$ and therefore the 
survival probability decays asymptotically as $t^{-\frac{v}{2n}}$, 
for $v=2mn$. On the other hand, since the coefficient of 
$z^{t}$ (for large $t$) in the expansion of 
$(1-z)^{\alpha}$ is proportional to $t^{-\delta}$, 
with $\delta = 1 + \alpha$, we conclude that the 
survival probability decays asymptotically 
as $t^{-v/2n}$, for $v \neq 2m\,n$ as well. Thus, we have 
$S(t) \sim t^{-\frac{v}{2n}}$ for arbitrary step lengths $v$ and 
$n$, which is the result we set out to prove.

\section{Numerical results} 
\label{sec/numerical} 
In this section we report exact numerical results for finite times   
($t \leq 10^4$) from iteration of the discrete time evolution equations. 
Consider first the hostile enviroment.  
Iteration of Eq. (\ref{evolution}), subject to the 
boundary conditions of Eqs. (\ref{bc at x=0}) and (\ref{bc at x=y}), 
yields the survival probability, $S(t)$, as shown in Fig. 2.  
Evidently, $S(t)$ approaches the asymptotic 
value, $2/t$, at long times.  It is interesting to examine the mode of 
approach to this scaling limit; assuming a power-law  
correction to scaling term, we write 
\begin{equation} 
S(t) \simeq \frac{2}{t}\left( 1 + \frac{A}{t^\phi} \right) ,
\label{cts} 
\end{equation} 
so that the dominant correction to scaling $\sim t^{-(1+\phi)}$. 
If this form is correct, then at long times 
$\ln \ln [tS(t)/2] \sim C - \phi \ln t$, where $C$ is a constant.  
Our results confirm the assumed correction to scaling
and yield an
exponent of $\phi \!=\! 1$ (see Fig. 2, inset). 
 
We have also analyzed, via iteration, the step-length combinations 
listed in Table \ref{tab1}.  

In all cases, the predicted value of $\delta$ is confirmed, and the 
correction to scaling exponent $\phi$ is unity. We have also verified numerically that, 
in all the cases studied, the mean position, conditioned on survival 
$\langle x \rangle_s \sim t^{1/2}$ and that  
$\langle x^2 \rangle_s \sim t$, as is to be expected. 
 
\section{Discussion} 
\label{sec/discussion}
We have studied the asymptotic survival probability 
of a random walker on the one-dimensional lattice, with the origin 
absorbing, and with a step-length that depends on whether the 
target site lies within the region that has been visited before. 
In all cases studied, we find that the survival probability decays 
asymptotically as a power law, $S(t) \sim t^{-\delta}$, 
where $\delta = v/2n$. 
Our expression for the decay exponent is 
in agreement with results obtained via numerical iteration of 
the transition matrix. 

\acknowledgments
We thank Miguel A. Mu\~{n}oz for helpful comments,   
in particular for suggesting the study of the     
friendly enviroment.   
We also thank Deepak Dhar for helpful discussions. 
R.D. and F.F.A. acknowledge financial support from CNPq (Brazil); 
D. b-A. acknowledges support of
NSF (USA) under grant PHY-0140094.

\newpage
\begin{table}[h]
\begin{center}
\begin{tabular}{|c|c|c|}
\hline
v       &       n       &       $\delta$\\
\hline
1       &       2       &       1/4\\
2       &       3       &       1/3\\
3       &       2       &       3/4\\
3       &       1       &       3/2\\
4       &       1       &       2\\
\hline
\end{tabular}
\caption{Step lengths studied via iteration}
\label{tab1}
\end{center}
\end{table}

\newpage 
\begin{center}  
\large{Figure Captions}  
\end{center}  
\vspace{1em} 
 
\noindent Fig. 1. Random walk in a hostile enviroment: transitions in the $x$-$y$ plane. 
\vspace{1em} 
  
\noindent Fig. 2. Main graph:  
decay of survival probability in the hostile model with $v\!=\!2$ and $n\!=\!1$; 
the equation of the solid line is $S = 2/t$. 
Inset: $\Delta = \ln[\ln(2/t)-\ln S(t)]$ versus $\ln t$; the slope of the 
straight line is $\phi=-1$. 
\vspace{1em}


\begin{thebibliography}{99}  

\bibitem{Barber}  
M. N. Barber and B. W. Ninham,   
\textit{Random and Restricted Walks},   
Gordon and Breach, New York, 1970.  
  

\bibitem{Bell}
K. De'Bell and T. Lookman, 
Rev. Mod. Phys. \textbf{65}, 87 (1993).

  
\bibitem{Marro/Dickman}  
J. Marro and R. Dickman,   
\textit{Nonequilibrium Phase Transitions in Lattice Models},   
Cambridge University Press, Cambridge, 1999.  
  
  
\bibitem{GCR}  
P. Grassberger, H. Chat\'{e}, G. Rousseau,   
Phys. Rev. E \textbf{55}, 2488 (1997).  
  
  
\bibitem{Jensen}  
I. Jensen, Phys. Rev. Lett. \textbf{70}, 1465 (1993);  
I. Jensen and R. Dickman, Phys. Rev. E \textbf{48}, 1710 (1993).  
 
 
\bibitem{Munoz/JSP} 
     M. A. Mu\~noz, G. Grinstein, and R. Dickman,  
     J. Stat. Phys. {\bf 91}, 541 (1998). 
      

\bibitem{Munoz/PRL} 
     M. A. Mu\~noz, G. Grinstein, R. Dickman, and R. Livi,  
     Phys. Rev. Lett. {\bf 76}, 451 (1996). 
      
     
\bibitem{Bohr}  
     T. Bohr, M. van Hecke, R. Mikkelsen, and M. Ipsen, 
     Phys. Rev. Lett. {\bf 86}, 5482 (2001). 
  
 
\bibitem{Dickman/ben-Avraham}  
R. Dickman and D. ben-Avraham,   
Phys. Rev. E \textbf{64}, 020102 (2001).  

  
\bibitem{van Kampen}  
N. G. van Kampen,   
\textit{Stochastic Processes in Physics and Chemistry},   
North-Holland, Amsterdam, 1992.  
 
  
\bibitem{Montroll/1965}  
E. W. Montroll, and G. H. Weiss,   
J. Math. Phys. \textbf{6}, 167-181 (1965).  
  
  
\end{thebibliography}
\end{document}